\documentclass[aps,prb,twocolumn,amsmath,amssymb,nofootinbib,superscriptaddress]{revtex4-2}

\usepackage{amsmath}
\usepackage{amssymb}
\usepackage{graphicx}
\usepackage{array}
\usepackage{bm}
\usepackage{braket}
\usepackage{booktabs}
\usepackage{dsfont}
\usepackage{dcolumn}
\usepackage{esint}
\usepackage{epstopdf}
\usepackage{mathrsfs}
\usepackage{tabu}
\usepackage{enumitem}

\newcommand{\bt}{\textbf}

\newcommand{\appropto}{\mathrel{\vcenter{
  \offinterlineskip\halign{\hfil$##$\cr
    \propto\cr\noalign{\kern2pt}\sim\cr\noalign{\kern-2pt}}}}}

\usepackage[colorlinks=true,citecolor=green,linkcolor=blue]{hyperref}
\usepackage{color}

\begin{document}

\title{Anomalous negative magnetoresistance in quantum-dot Josephson junctions with Kondo correlations}
\author{M.-T.~Deng}
\thanks{These authors contributed equally to this work.}
\affiliation{Institute for Quantum Information \& State Key Laboratory of High Performance Computing, College of Computer Science and Technology, NUDT, Changsha 410073, China}
\affiliation{NanoLund and Division of Solid State Physics, Lund University, Lund S-22100, Sweden}
\affiliation{Hefei National Laboratory, Hefei 230088, China}

\author{C.-L.~Yu}
\thanks{These authors contributed equally to this work.}
\affiliation{China Greatwall Quantum Laboratory, Changsha 410006, China}
\affiliation{NanoLund and Division of Solid State Physics, Lund University, Lund S-22100, Sweden}

\author{G.-Y.~Huang}
\email[Corresponding author: ]{guangyaohuang@quanta.org.cn}
\affiliation{Institute for Quantum Information \& State Key Laboratory of High Performance Computing, College of Computer Science and Technology, NUDT, Changsha 410073, China}

\author{R.~L\'{o}pez}
\affiliation{Institut de F\'{\i}sica Interdisciplin\`aria i de Sistemes Complexos IFISC, Palma de Mallorca E-07122, Spain}

\author{P.~Caroff}
\affiliation{NanoLund and Division of Solid State Physics, Lund University, Lund S-22100, Sweden}

\author{S.~Ghalamestani}
\affiliation{NanoLund and Division of Solid State Physics, Lund University, Lund S-22100, Sweden}

\author{G.~Platero}
\affiliation{Instituto de Ciencia de Materiales de Madrid, Madrid 28049, Spain}

\author{H.Q.~Xu}
\email[Corresponding author: ]{hqxu@pku.edu.cn}
\affiliation{NanoLund and Division of Solid State Physics, Lund University, Box 118, S-221 00 Lund, Sweden}
\affiliation{Beijing Key Laboratory of Quantum Devices, Key Laboratory for the Physics and Chemistry of Nanodevices, and Department of Electronics, Peking University, Beijing 100871, China}
\affiliation{Beijing Academy of Quantum Information Sciences, Beijing 100193, China} 

\begin{abstract}
The interplay between superconductivity and the Kondo effect has stimulated significant interest in condensed matter physics. They compete when their critical temperatures are close and can give rise to a quantum phase transition that can mimic Majorana zero modes. Here, we have fabricated and measured Al-InSb nanowire quantum dot-Al devices. In the Kondo regime, a supercurrent-induced zero-bias conductance peak emerges. This zero-bias peak shows an anomalous negative magnetoresistance (NMR) at weak magnetic fields. We attribute this anomalous NMR to quasiparticle trapping at vortices in the superconductor leads as a weak magnetic field is applied. The trapping effect lowers the quasiparticle-caused dissipation and thus enhances the Josephson current. This work connects the vortex physics and the supercurrent tunneling in Kondo regimes and can help further understand the physics of Josephson quantum dot system.
\end{abstract}

\date{\today}

\maketitle

External magnetic fields generally suppress superconductivity by the orbital effect~\cite{Gennes1964} and the Zeeman effect~\cite{Gennes1989}. Nevertheless, the magnetic field-induced or -enhanced superconductivity was proposed theoretically~\cite{Jaccarino1962, Maekawa1978} and observed experimentally in many superconductors~\cite{Meul1984, Uji2001, Kogan1987}. Most of these counterintuitive field-enhanced superconductivities are related to the field compensation to internal magnetic moments of the superconductors (such as the Jaccarino-Peter effect) or to the field suppression of Kondo spin-flipping induced Cooper-pair breaking~\cite{Abrikosov1969, Maki1969}. The latter mechanism is crucial for the superconductors that are doped with magnetic impurities.

\begin{figure*}
\centering
\includegraphics[width=17 cm]{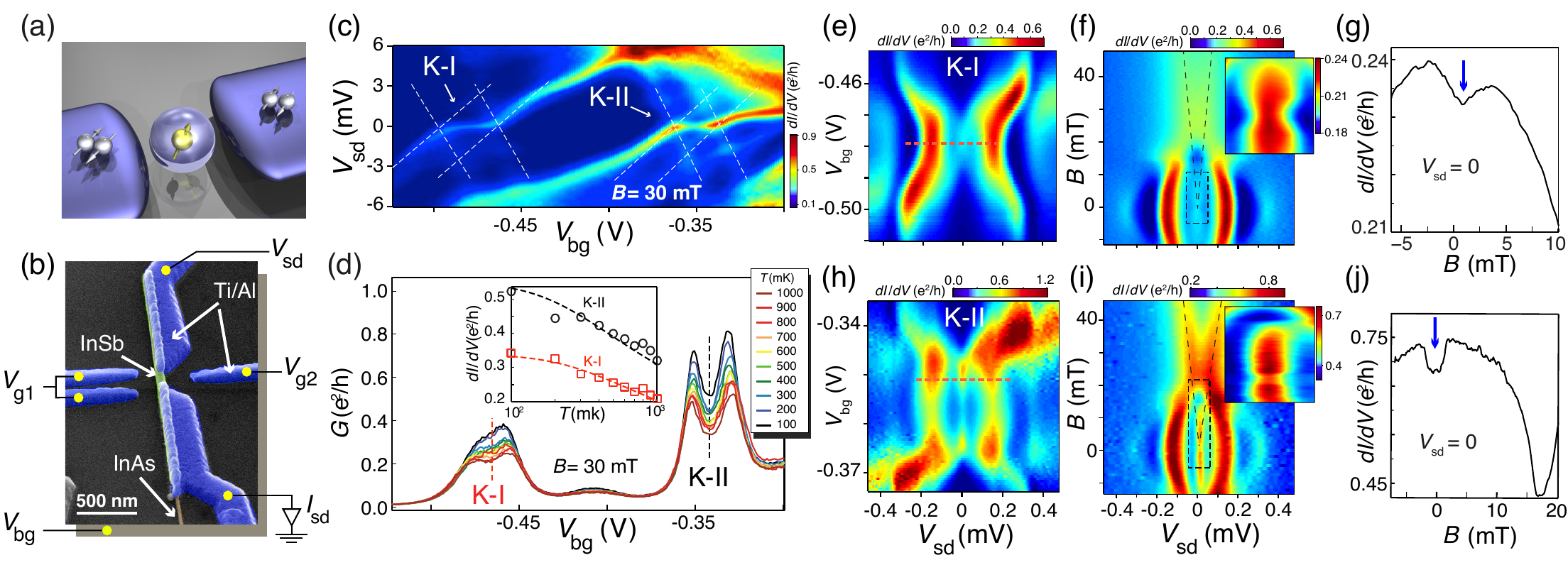}
\caption{ (Color online) Al-InSb nanowire-Al Josephson quantum dot device. (a) Conceptional sketch of a typical Josephson quantum dot system. (b) Scanning electron microscope (SEM) image of Device~1. (c) Differential conductance as a function of $V_{\rm sd}$ and $V_{\rm bg}$ at $B=30$~mT. Zero-bias conductance ridges (marked by K-I and K-II) arising from the Kondo effect can be seen in two odd-occupation regions.  (d) Conductance of the device as a function of $V_{\rm bg}$ measured in the linear response regime ($V_{\rm sd} = 10~\mu$V) at $B=30$~mT and at different temperatures. Inset: Conductance extracted as a function of temperature at the K-I (red squares) and K-II (black circles) Kondo anomaly valleys and theoretical fits (dashed lines) to the data~\cite{Gordon1998}. (e) Low energy conductance spectrum measured for the device in the superconducting state ($B=0$) in the K-I valley as a function of $V_{\rm bg}$. (f) Conductance spectrum as a function of magnetic field at ${V_{\rm bg}}= -480$ mV [indicated by the red dashed line in panel (e)]. Inset: Close-up view of the measurements in the dashed square region. (g) Line-cut at $V_{\rm sd}=0$ taken from (f). Indicated by the blue arrow, there is a conductance valley at zero and low magnetic fields. (h)-(j) Similar to (e)-(g), but measured in the K-II valley. Here, an even more pronounced NMR at low fields is seen. All data in this figure are taken from Device~1 with $V_{\rm g1}=V_{\rm g2}=0$. In the K-I (K-II) valley, the charging energy $U$ is about 2.5 (2.3) meV, the electron g-factor $g^* \approx48$ (52), and the coupling asymmetry $\Gamma_L/\Gamma_R\approx0.04$ (0.09). The superconducting gap of the leads is $\Delta\approx0.16$ meV.} 
\label{Fig1}
\end{figure*}

The rich correlation physics of a magnetic impurity and a superconductor could also be studied by employing a superconductor-quantum dot-superconductor (quantum-dot Josephson junction) device. In such a device, the quantum dot can be tuned to a spinful ground state [Fig.~\ref{Fig1}(a)] and, with an unpaired spin inside, is an analog of a single magnetic impurity, yet highly manipulable. 

In this Article, we report on a study of Cooper-pair transport through quantum dot Josephson junction devices in the spinful Coulomb blockade regime. In the presence of spin correlation to Cooper pairs, a Kondo resonance-enhanced supercurrent emerges as a zero-bias peak (ZBP). We observe that the ZBP exhibits a pronounced suppression in the vicinity of zero magnetic field, namely that a steep dip appears on the ZBP around zero field, and the ZBP recovers to a maximum quickly with increasing magnetic field. We ascribe this observed anomalous negative magnetoresistance (NMR) to the enhancement of Josephson switching current due to quasiparticle trapping at vortices. 

We study three Al-InSb nanowire-Al quantum dot Josephson junction devices, denoted as Device~1--3, respectively. In each device, a quantum dot is formed in an InSb nanowire segment between two Al leads~\cite{Nilsson2009} due to the combined effect of native or induced charges, electrical field screening of metal, and the surface or interface band bending~\cite{Kretinin2010}. The quantum states of the dots and the dot-lead couplings can be tuned by the back gate voltage ($V_{\rm bg}$) and the side gate voltages ($V_{\rm g1}$ and $V_{\rm g2}$) [Fig.~\ref{Fig1}(b)]. The measurements are performed in a wet \textsuperscript{3}He/\textsuperscript{4}He dilution refrigerator. Unless specified, all data are taken at the cryogenic base temperature ($\sim$25 mK), and the external magnetic fields are all applied perpendicularly to the device substrates. The fabrication and measurement details can be found in the Methods.

We first characterize the devices in the normal state at magnetic field $B=30$~mT, where the superconductivity is quenched. As shown in Fig.~\ref{Fig1}(c), the normal state differential conductance of Device~1, $dI/dV$, measured as a function of source-drain bias voltage $V_{\rm sd}$ and back-gate voltage $V_{\rm bg}$, displays typical lifetime broadened Coulomb blockade diamond structures. Two horizontal high conductance anomalies at zero-bias voltage emerge inside the diamonds at $V_{\rm bg}\approx -0.5$~V to $-0.46$~V (denoted as the K-I ridge) and $V_{\rm bg}\approx -0.37$~V to $-0.33$~V (denoted as the K-II ridge), which can be attributed to the Kondo effect. The temperature dependence of the two anomalies shown in Fig.~\ref{Fig1}(d) also display a typical Kondo behavior, i.e., the zero-bias anomalies drop as the temperature increases. Utilizing the relation, $G=G_{0}\left ( \frac{{T_K^{'}}^{2}}{T^{2}+{{T}^{'}_K}^{2}}\right )^s$, where $T^{'}_K=\frac{T_K}{{\left ( 2^{1/s}-1 \right )}^{1/2}}$ and $s=0.22$~\cite{Gordon1998}, we can estimate the normal state Kondo temperature $T_K$, yielding $T_K\approx 1.6$~K at the middle of the K-I ridge and $T_K\approx 1.2$~K at the middle of the K-II ridge. 

\begin{figure}[t]
\centering
\includegraphics[width=8cm]{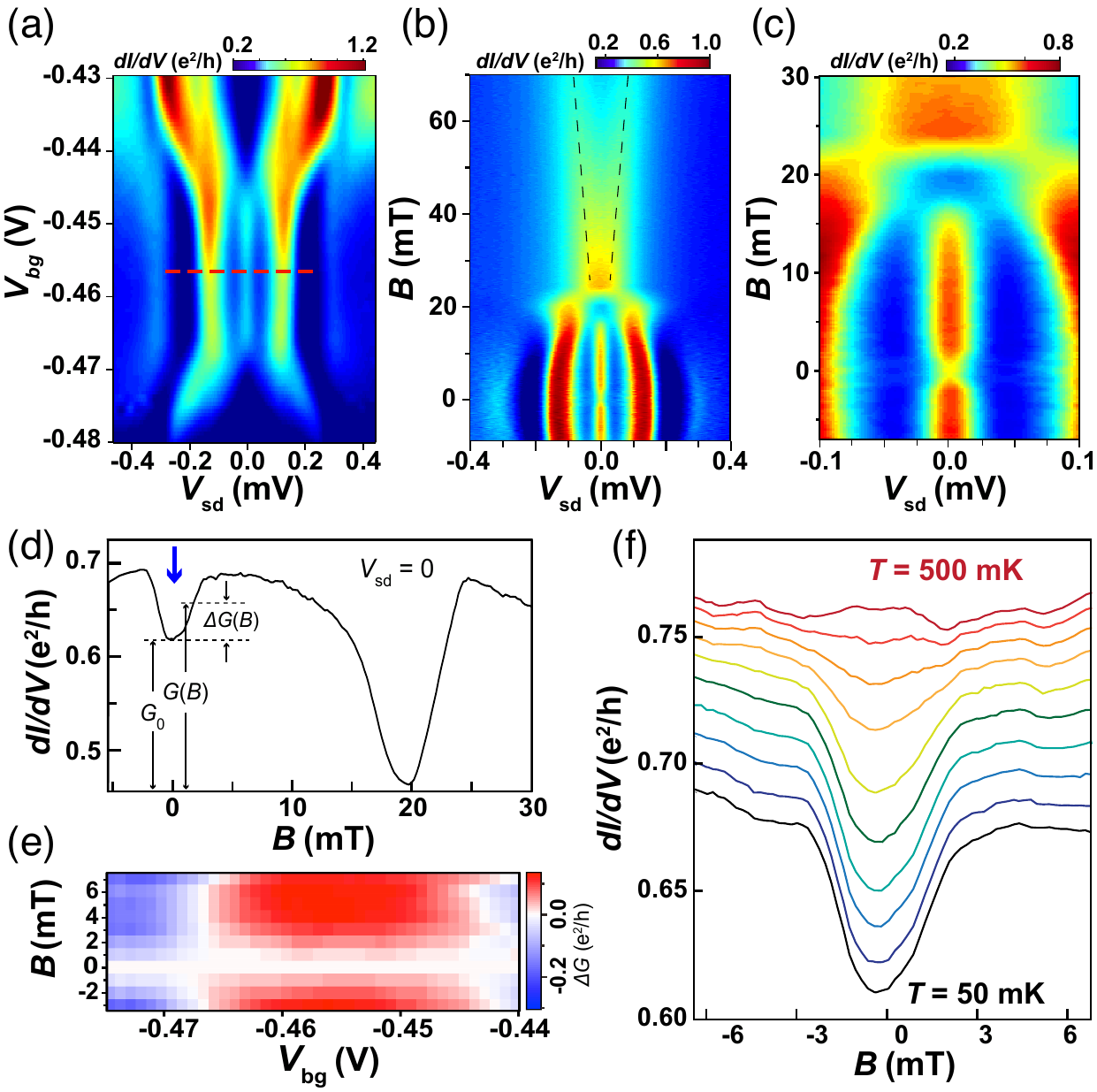}
\caption{(Color online) Anomalous NMR in Device~2. (a) Low energy conductance spectrum in the superconducting state ($B=0$) in a Kondo-correlated Coulomb blockade region as a function of $V_{\rm bg}$. (b) Magnetic field evolution of low energy conductance spectrum at $V_{\rm bg}=-457$~mV [the red dashed line in (a)]. (c) Close-up view of the measurements in the low-bias voltage regime of panel (b). (d) Line-cut at $V_{\rm sd}=0$, taken from panel (c), along the $B$-axis. Similar to Device~1, a sharp conductance dip on the ZBP is seen in (b)-(d). (e) The extracted $\Delta G(B)$ [see definition in (d)] values as a function of $V_{\rm bg}$ and $B$. (f) Temperature dependence of the low-field NMR dip. The NMR feature is gradually smeared as the temperature increases and is completely washed out at $T\sim 450$~mK. The line-cuts are successively offset by 0.01 $e^2/h$ for clarity. All data are taken from Device~2 at $V_{\rm g1}=-3.5$~V and $V_{\rm g2}=3$~V. Related parameters: $U = 2.1$ meV, $g^* \approx41$, $T_K\approx 1.4$ K (estimated at 30 mT), $\Delta\approx0.16$ meV, and $\Gamma_L/\Gamma_R\approx0.08$.}
\label{Fig2}
\end{figure}

The conductance spectrum is significantly modulated when the leads are in the superconductor state. As shown in Figs.~\ref{Fig1}(e) and \ref{Fig1}(h), at zero magnetic fields, the conductance spectra corresponding to the K-I and K-II ridges show two high conductance stripes at finite bias bounded by $V_{\rm sd}=\pm2\Delta/e$ (where $\Delta$ is the superconductor gap). The external magnetic field drives these sub-gap structures to lower energies [Figs.~\ref{Fig1}(f) and \ref{Fig1}(i)] until $B\approx 25$~mT, at which a recovered Kondo resonance is observable. These sub-gap structures can be ascribed to multiple Andreev reflections or Andreev bound states (ABSs)~\cite{Kim2013, Li2017, Buitelaar2002, Buitelaar2003, Jespersen2007, Eichler2007, Pillet2010, Lee2014}. Already covered by the above-cited landmark papers, the sub-gap structures will not be further discussed in detail in this work. 

In the presence of charging energy, Cooper pair tunneling through the quantum dot is suppressed in the Coulomb blockade regime. When there is a strong Kondo correlation, however, an induced density of states (DOS) peak pinned at the Fermi level of the dot can significantly enhance the conductance of the quantum dot Josephson junction and allow Cooper-pair tunneling~\cite{Rasmussen2007}. We have observed such Cooper-pair tunneling-induced ZBPs in both the K-I and the K-II Coulomb blockade regions. Especially in the K-II valley, the ZBP is seen to run through the entire blockade region [see Fig.~\ref{Fig1}(h)]. 

Surprisingly, in the magnetic field evolution measurements of the ZBPs, a dip structure, i.e., an NMR characteristic, is seen to develop at low magnetic fields [see Figs.~\ref{Fig1}(f)-\ref{Fig1}(g) and \ref{Fig1}(i)-\ref{Fig1}(j)]. The height of the ZBP increases from $B=0$ to $B\approx3$ mT and then decreases as the magnetic field further increases. Especially in the K-II region, the ZBP shows a steep NMR with a $\sim$0.08 $e^2/h$ conductance enhancement from $B=0$ to $B\approx3$ mT.

This anomalous low-field NMR has also been observed in Device~2 and Device~3. Device~2 is lithographically identical to Device~1, while Device~3 has slimmer contacts than Devices~1 and 2. As shown in Fig.~\ref{Fig2}(a), a ZBP is developed in a Kondo-correlated Coulomb blockade region of Device~2. Similar to Device~1, in the magnetic field evolution measurements of the ZBP, a steep conductance dip in the range of $B=-3$~mT to 3~mT is clearly observed [Figs.~\ref{Fig2}(b)-\ref{Fig2}(d)]. We define $\Delta G (B)$ as the difference between the zero-bias conductance $G(B)$ at a magnetic field $B$ and the zero-field conductance $G_0$ [see the definition in Fig.~\ref{Fig2}(d)]. We perform the same gate dependence measurements at various magnetic fields and extract $\Delta G$ as a function of gate voltage and magnetic field. The extracted results are plotted in Fig.~\ref{Fig2}(e), and it is seen that the NMR appears in the entire gate voltage range. We also note that the NMR width is almost constant in the whole $V_{\rm bg}$ range, extending from $\sim -3$ mT to $\sim 3$ mT. This anomalous, sharp NMR and the $V_{\rm bg}$ independence are also observed at other gate voltages. More examples are shown in Fig.~\ref{FigAPP_Dev2_1} in Appendix~\ref{sec:level5}.

We have also performed temperature-dependence measurements for the anomalous NMR at low fields [Fig.~\ref{Fig2}(f)]. The anomalous NMR dip is seen to gradually smear out as the temperature increases and vanish completely at $T\sim 450$~mK which is well below the critical temperature of the Al leads ($\approx1.2$~K). 

\begin{figure*}[t]
\centering
\includegraphics[width=16 cm]{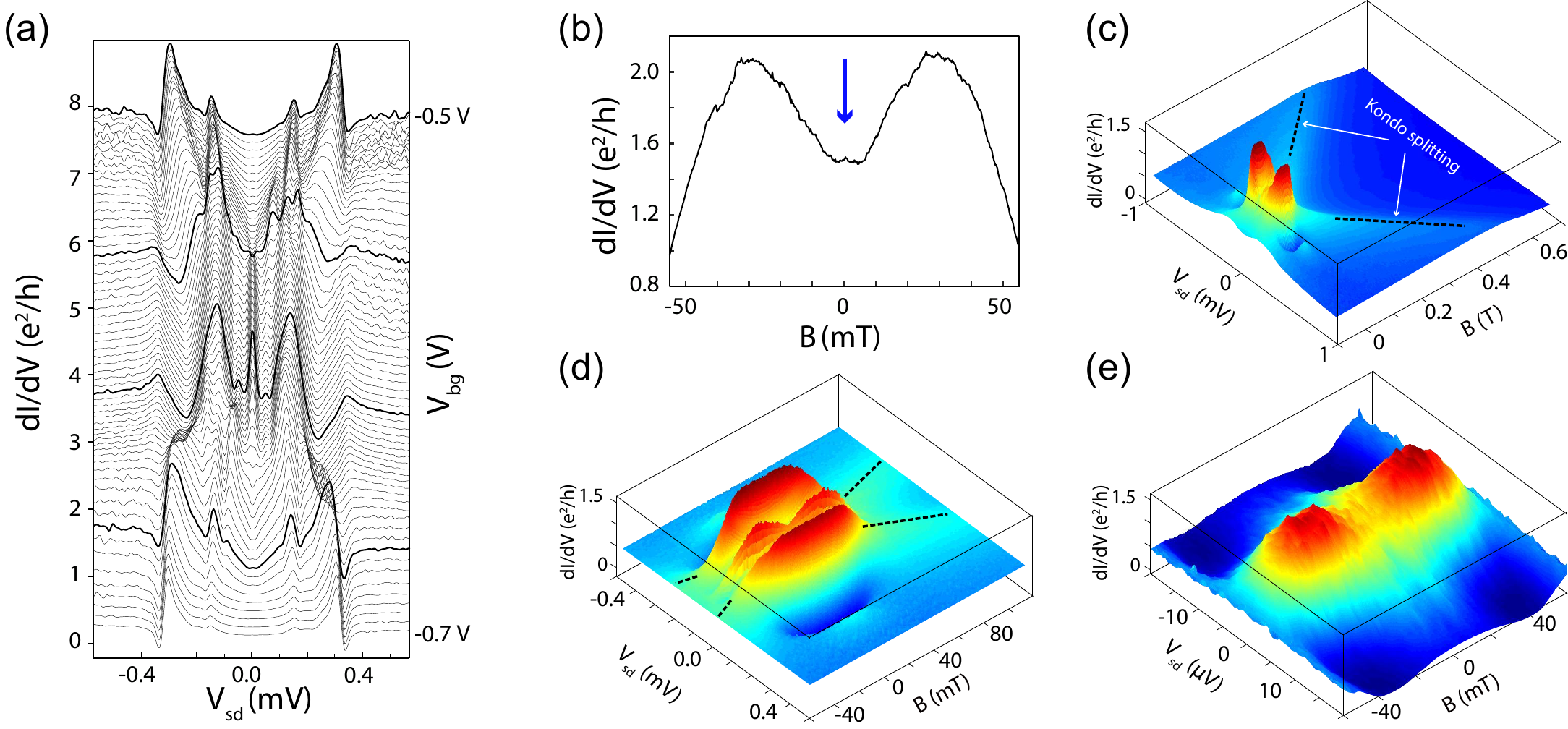}
\caption{\label{Fig3} (Color online) Anomalous NMR in Device~3. (a) Line-cut plot of the conductance spectra of Device 3 in a Kondo-correlated Coulomb blockade region at $B=0$ (in the superconducting state), where Sub-gap conductance peaks and ZBPs are clearly visible. Traces are successively offset by 0.12~$e^2/h$. (b) NMR structure on the ZBP measured for the device at $V_{\rm bg}=-0.585$ V (i.e., at the middle of the blockade region). The depth of the NMR dip is about 0.6~$e^2/h$.   (c)-(e) 3D color plots of the differential conductance of the device measured as a function of $V_{\rm sd}$ and $B$ at $V_{\rm bg}=-0.585$ V, where the magnetic field evolution of the ZBP can be viewed by taking the line cut at $B=0$. Panel (c) shows the measurements in a large range of $B$ and $V_{\rm sd}$. Panel (d) shows the measurements taken in much higher resolutions of $B$ and $V_{\rm sd}$ than in panel (c). Panel (e) shows the measurements taken in even higher resolutions of $B$ and $V_{\rm sd}$. Other characteristic values extracted for the device in the measured gate voltage range are $U \approx7.5$ meV, $g^* \approx40$, $\Delta=0.16$ meV, and $T_K\approx2.4$ K (at $B=70$ mT).}
\end{figure*}

The devices are also characterized in the open regime (where the Coulomb blockade is absent). As shown in Fig.~\ref{FigAPP_Dev2_2} in Appendix~\ref{sec:level4}, the supercurrent-induced ZBPs in the open regimes are even more pronounced due to the absence of tunneling barriers. In contrast to the anomalous NMR observed in the Kondo regime, no NMR is observed in the open regime. 

A similar anomalous NMR is also observed in Device~3. As shown in Fig.~\ref{Fig3} (and Figs.~\ref{FigAPP_Dev3_1} and~\ref{FigAPP_Dev3_2} in Appendix~\ref{sec:level6}), the supercurrent induced ZBP emerges in the Kondo regime of the device. In the magnetic field evolution, the ZBP shows a pronounced NMR dip at the low-field region, whose depth is up to 0.6 $e^2/h$. However, the width of NMR is much larger than those observed in Devices~1 and 2. For Device~3, the maximum ZBP height appears around $\pm 30$~mT, almost an order of magnitude higher than the NMR dip widths of Devices~1 and 2. We will discuss the reason of the width difference later.


We now discuss the origin of the anomalous NMR. In the superconductivity-Kondo interplay regime, several possible mechanisms could give rise to an NMR. For example, the magnetic field could increase the $k_{\rm B}T_{\rm K}/\Delta$ ratio and, therefore, induce a recovered Kondo DOS at the Fermi level~\cite{Lee2012}. However, the NMRs observed in our study occur well below the critical field of the superconductor Al leads, at which $k_{\rm B}T_{\rm K}/\Delta$ has not yet been effectively increased. Actually, as shown in Figs.~\ref{Fig1}(f), \ref{Fig1}(i) and \ref{Fig2}(c), we do observe the existence of a recovered Kondo DOS, however, at a significantly higher field and has a very different magnetic-field dependent profile. Due to a similar reason, it is also unlikely that the observed NMRs are related to the field-induced topological phase transition~\cite{Mourik2012, Deng2012, Deng2016, Tiira2017, Fu2021, Cao2023} or weak antilocalization physics~\cite{Pikulin2012}. 

Another possible reason for an NMR is the magnetic field-induced quantum phase transition (QPT). Suppose a Josephson quantum dot is initially at a ``0-$\pi$" phase crossing point at zero field. In that case, applying a magnetic field can drive the device away from the crossing point and increase the (reversal) Josephson current~\cite{vandam2006}. This QPT will also give rise to an emergent zero-energy ABS~\cite{Lee2014, Chang2013}. However, the observed NMRs here appear in an extensive range of $T_K$ and thus are less likely due to the QPT at the ``0-$\pi$" phase crossing point. This is true even when the presence of the $0^{\prime}$ and $\pi^{\prime}$ phases of the quantum dot Josephson junctions \cite{Lee2022} are considered. In fact, the curvature of the ABS conductance in Fig.~\ref{FigAPP_Dev2_1}(a) in Appendix~\ref{sec:level5} (the dashed line) indicates that the quantum dot is in a well-defined spin-doublet ground state in the middle of the Kondo-correlated Coulomb blockade region~\cite{Kim2013}. Hence, there is no field-induced QPT in this region, and the QPT could not be the physical origin of the observed anomalous NMRs in this work. 

\begin{figure}[t]
\centering
\includegraphics[width=8.5 cm]{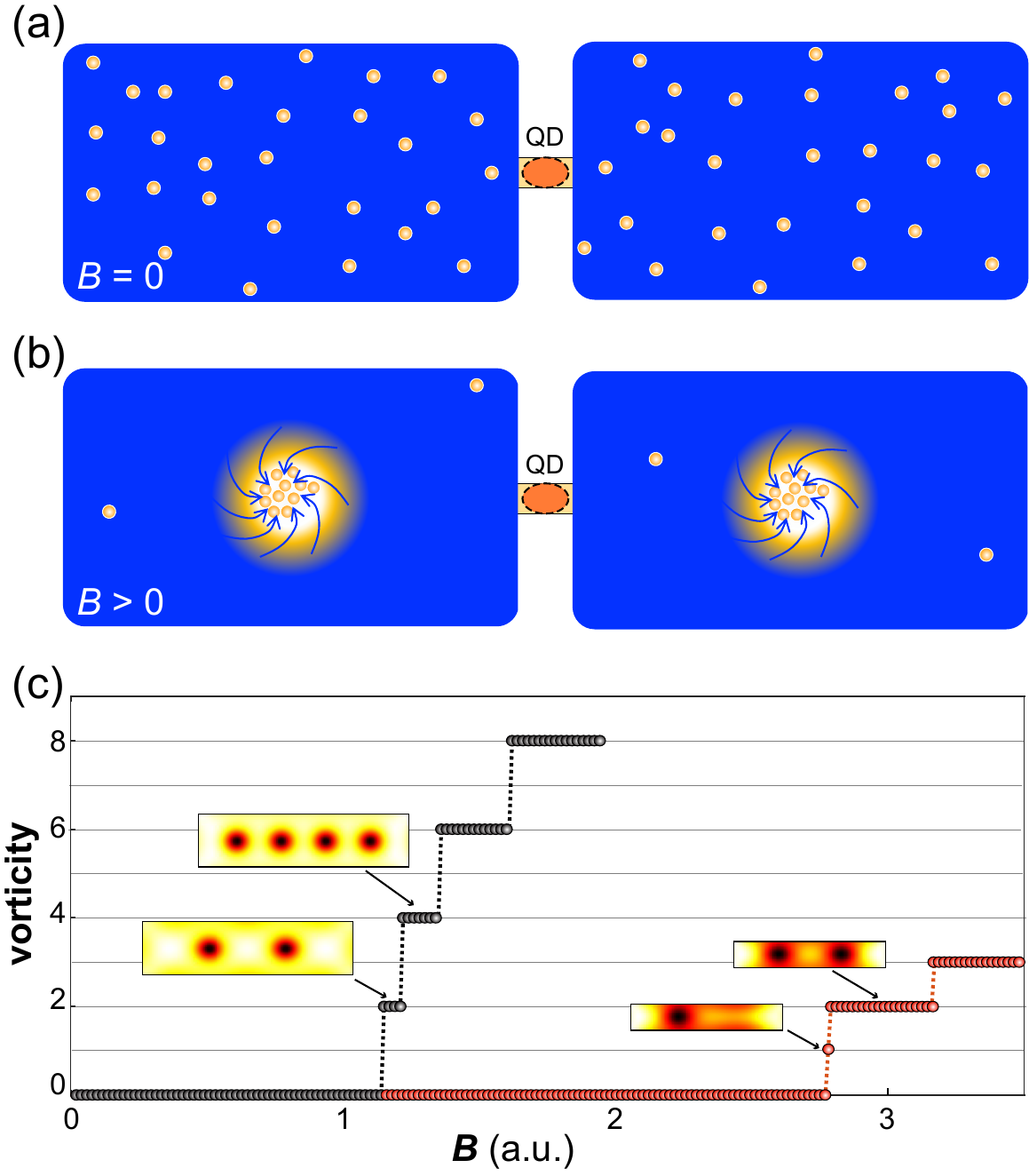}
\caption{ (Color online) Vortices as quasiparticle traps. (a)-(b) Schematics of a nanowire quantum-dot Josephson junction at zero [panel (a)] and finite [panel (b)] magnetic field. Small yellow dots represent quasiparticles in the leads. Vortices formed in the superconductor leads at finite magnetic field as shown in panel (b) can serve as quasiparticle traps. (c) Simulation of the superconductor vorticity based on the Ginzburg-Landau theory as a function of applied magnetic field. The black and red circles denote data for wide and narrow leads, respectively. For the same superconductor material, the magnetic field onset for vortex formation is higher for the device with narrower leads.}
\label{Fig4}
\end{figure}

To analyze the cause of the anomalous NMR, it is worth emphasizing that the NMR width is seemingly independent of the gate voltage (thus the Kondo temperature) or the $g$-factor of the quantum dot.  Nevertheless, the NMR width does show a dependence on device structure. For Devices~1 and~2, which have similar geometry, the NMR width is about 6 mT. In contrast, the NMR width for Device~3, with narrower contacts, is almost 60 mT (see Fig.~\ref{Fig3}). Therefore, we attribute the observed NMR of the ZBP in Kondo regimes to the enhancement of the switching supercurrent ($I_{\rm sw}$) by quasiparticle trapping at vortices. As mentioned above, the ZBP formed in the Kondo valley is due to the cotunneling of Cooper pairs, i.e., Josephson current. Usually, substantial quasiparticles exist in the superconductor leads even at low temperatures~\cite{Wang2014NC}. These quasiparticles can tunnel through the Josephson junction, cause dissipation, and switch the tunneling current from a dissipationless current to a dissipative one~\cite{Ambegaokar1969}. As a weak magnetic field is applied, vortices form in the superconductor thin film leads, diminishing the order parameter to zero in the cores. Hence, thermally excited quasiparticles can be trapped in vortices and recombine to Cooper pairs (Fig.~\ref{Fig4}). The quasiparticle dissipation effect and the effective electron temperature are therefore significantly suppressed due to the trapping effect, and the ZBP, as a metric of the $I_{\rm sw}$ of the junction, gets enhanced.

When the applied magnetic field is large, the ZBP will decrease again. The reason is two-fold. On the one hand, a large magnetic field can destruct Cooper pairs, like in most cases when a magnetic field is applied to a superconductor. On the other hand, the Zeeman effect splits the Kondo peak and reduces the electrical transparency of the junctions. 

The quasiparticle-trapping explanation is consistent with the independence of $T_K$ or $g$-factor of the NMR width since the vortices are formed in the leads and independent of the quantum dot properties. It is also consistent with the fact that the NMR width of Device~3 is quite different from Devices~1 and 2 because the formation of vortices strongly depends on the size and geometry of the superconductor contacts~\cite{Pekola2013RMP, Stan2004PRL}.  In Fig.~\ref{Fig4}(c), we have performed a numerical study on the formation of vortices based on the Ginzburg-Landau theory. It is evident that for the same superconductor material, narrower contacts need higher magnetic fields to form vortices than wider contacts.

\begin{figure}[h]
\centering
\includegraphics[width=8.5 cm]{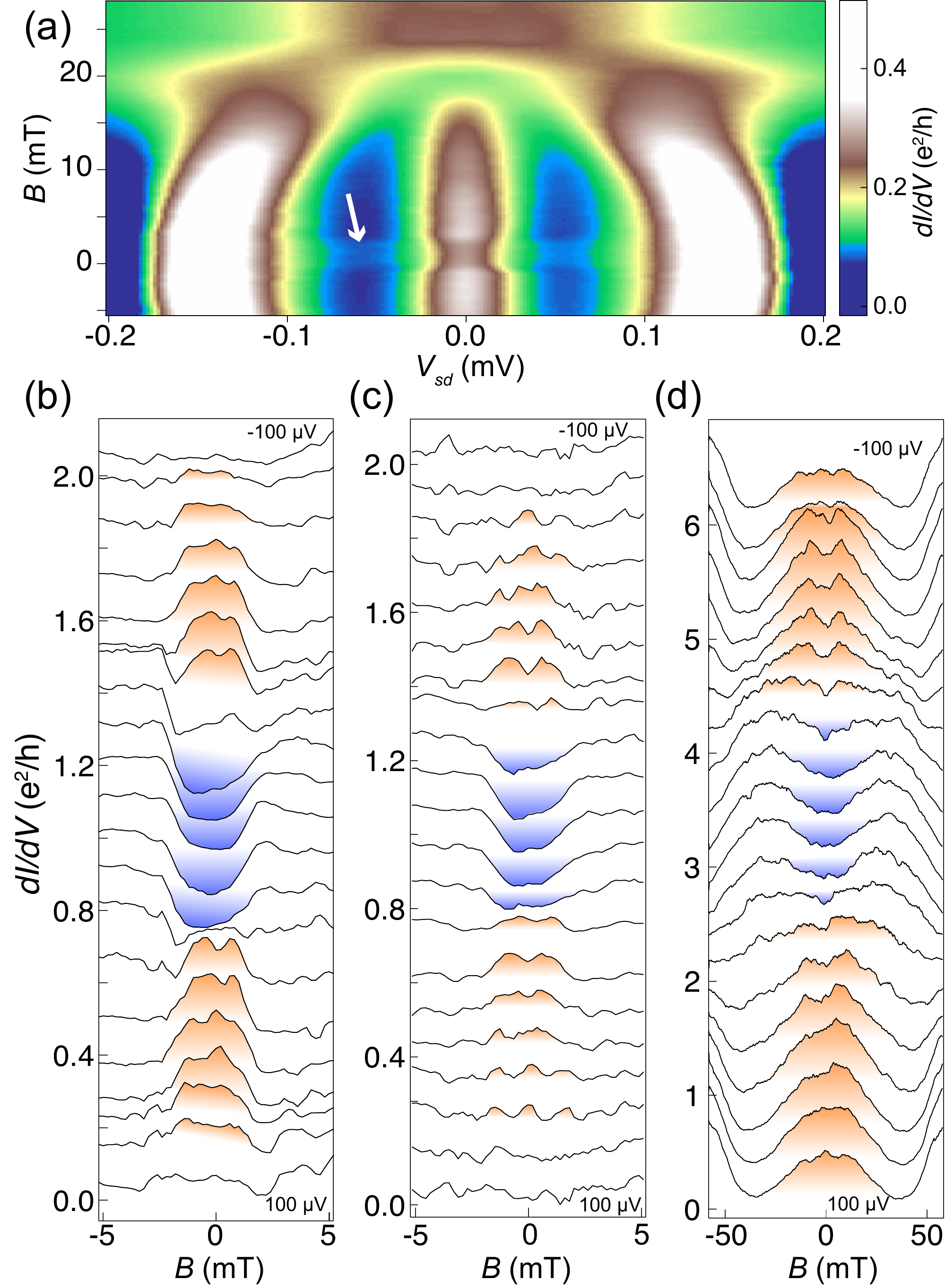}
\caption{ (Color online) Conductance bumps at finite bias voltages. (a) Differential conductance of Device 2 as a function of $V_{\rm sd}$ and $B$, measured at $V_{\rm bg}=-0.436$ V, $V_{\rm g1}$=1 V and $V_{\rm g2}=-1$ V. A conductance bump (indicated by a white arrow) is developed at finite bias voltages. (b) Line-cut view of the low-field regime of panel (a). A transition from conductance bump (brown) at high bias voltages to conductance dip (blue) at low bias voltages is clearly shown. (c)-(d), Similar to panel (b), but for line-cuts obtained form Fig.~\ref{Fig2}(c) and Fig.~\ref{Fig3}(e), respectively. The lines in panels (b)-(d) are sequentially offset for clarity.}
\label{Fig5}
\end{figure}

To examine the quasiparticle-trapping picture, we have further studied the tunneling conductance at finite bias in the Kondo regime. As shown in Figs.~\ref{Fig5}(a) and (b), there is a conductance bump at finite bias voltages between $B\approx-3$ mT and $B\approx3$ mT, coinciding with the NMR range. The conductance bump can be ascribed to the quasiparticle tunneling at finite bias at a low magnetic field. Once the quasiparticles are trapped into vortices at a higher magnetic field, the tunneling current drops due to the lack of charge carriers within the superconductor gap. In other words, the conductance bumps and the ZBP-NMR are both caused by quasiparticle trapping at vortices. Similar conductance bumps associated with the NMR can also be found in other regions of Device~2 [Fig.~\ref{Fig5}(c)] and the Kondo regime of Device~3 [Fig.~\ref{Fig5}(d)]. 

The quasiparticle trapping at vortices has also been observed in other literature, such as in superconductor qubit devices~\cite{Wang2014NC} and InAs nanowire Josephson junctions~\cite{Sato2022PRL}. Especially in Ref.~\cite{Sato2022PRL}, the $I_{\rm sw}$ enhancement due to the quasiparticle-trapping at vortices has been observed, but without a quantum dot or Kondo resonance being identified.

There are still two experimental facts that need to be addressed. Firstly, the NMR smears out at higher temperatures but still well below the critical temperature of the superconductor. This can be understood as the saturation of the trapping effect. The number of quasiparticles increases dramatically as the temperature increases and finally exceeds the trapping capacity of vortices~\cite{Wang2014NC}. The same temperature dependence has also been reported in Ref.~\cite{Sato2022PRL}. Actually, the increase of quasiparticle density in the devices can be identified in the tunneling spectra at different temperatures, see  Fig.~\ref{FigAPP_Dev1} in Appendix~\ref{sec:level4} and Fig.~\ref{FigAPP_Dev3_2} in Appendix~\ref{sec:level6} (also see Ref.~\cite{Li2017}). 

Secondly, the NMR does not appear in the open regime of the devices studied in this work (Fig.~\ref{FigAPP_Dev2_2} in Appendix~\ref{sec:level5}). This could be attributed to several mechanisms. The junction is more electronically transparent in the open regime, where the gate voltage is more positively biased. In the RSCJ model, this change will affect the damping of the junction and its sensitivity to quasiparticles. In addition to a change in  the junction transparency, the strength of the proximity effect and the charge density in the nanowire segments underneath the superconductor will also differ from the blockade regime. The proximity effect change may lead to a redistribution of vortices, while the charge density change may influence the quasiparticle number in the nanowire~\cite{Vaitiekenas2018PRL, Mikkelsen2018PRX, Antipov2018PRX, Loo2023}. All of the above possibilities, or part of them, could cause the absence of the NMR in the open regimes.

In summary, we have observed low-field NMRs on the ZBPs in quantum dot Josephson junctions in the Kondo-superconductivity interplay regime. The NMR occurs at Kondo regimes with various $T_K$ but shows a robust device-geometry dependence. We ascribe the anomalous NMR to the quasiparticle trapping at vortices in the superconductors in weak magnetic fields. This is confirmed by the emerged conductance bump at finite bias voltages in the same range as the NMR appears. The trapping effect is greatly weakened when the quasiparticle density increases with the temperature or when substantial quasiparticles are generated locally in the nanowires in open regimes. This observation shed light on the connections between the vortex physics and Kondo correlations in Josephson devices for the first time. It can help to design topological devices with mitigated quasiparticle-poisoning.

\bt{Acknowledgements.} This work was financially supported by the Swedish Research Council (VR), the National Natural Science Foundation of China (Nos.92165208, ~11874071, 91221202, 91421303, and 11904399), the Ministry of Science and Technology of China (MOST) through the National Key Research and Development Program of China (No.~2016YFA0300601 and 2017YFA0303304), and the Beijing Academy of Quantum Information Sciences (No.~Y18G22). M.T.D. acknowledges the financial supports by the Hunan Provincial Science Foundation for Distinguished Young Scholars (Grant No. 2021JJ10043) and the Innovation Program for Quantum Science and Technology (Grant No. 2021ZD0302401). R.L. acknowledges the financial support by  the Spanish Ministry of Economy and Competitiveness through Grants No. PID2020-117347GB-I00 and the María de Maeztu project CEX2021-001164-M funded by the MCIN/AEI/10.13039/501100011033. G.P. acknowledges the financial supports by the Spanish Ministry of Economy and Competitiveness through the grant:PID2020-11778GB-I00, and the supports from CSIC Research Platform PTI-001 and through the grant LINKB20072(CSIC).

\vspace{5pt}
\textbf{Contributions.} C.L.Y. fabricated the devices. M.T.D. and C.L.Y. performed the measurements. G.Y.H. performed the Ginzburg-Landau simulation of vortices.  P.C., and S.G.G. grew the semiconductor nanowires. M.T.D., C.L.Y., R.L., G.P., and H.Q.X. analyzed the data. H.Q.X. led the experiments and supervised the work. All of the authors contributed to the manuscript writing and discussion.

\begin{appendix}

\section{}
\subsection{\label{sec:level1} Details of device fabrication}
Each of the devices studied in the article constitutes an InSb nanowire and two Al superconductor leads. To fabricate these devices, high quality InAs/InSb (about 40/80~nm in diameter) heterostructure nanowires were grown by metal organic vapor phase epitaxy [Caroff, P. et al. \emph{Small} \textbf{4}, 878 (2008)] and some of these nanowires were transferred to a highly \emph{n}-doped Si/SiO\textsubscript{2} substrate. After the transfer, a few nanowires were selected and the InSb nanowire segment of each selected nanowire was then contacted with a pair of thermally evaporated titanium/aluminium (Ti/Al, 5/100 nm) leads with a separation of about 150~nm. To obtain a clean metal-semiconductor interface, the nanowire contact areas were etched/passivated in a diluted (NH\textsubscript{4})\textsubscript{2}S\textsubscript{x} solution before metal deposition [Nilsson, H.A. et al. \emph{Nano Lett.} \textbf{12}, 228 (2012)]. For Device~1 and Device~2 studied in the main article, Al side gates were also fabricated in the same step as for the source-drain contacts.

\subsection{\label{sec:level2} Measurement setup}
In the measurements, a source-drain bias voltage, $V_{\rm sd}$, was antisymmetrically applied to each device, i.e., with $-\frac{1}{2}V_{\rm sd}$ and $\frac{1}{2}V_{\rm sd}$ applied to the source and drain contacts, using a home-made DC voltage source with a 2~$\mu$V resolution. Current was measured through a home-made amplifier and it was then converted to differential conductance $dI/dV$ numerically. Gate voltages were applied from commercial DC voltage sources. All the measurements were performed in a \textsuperscript{3}He/\textsuperscript{4}He-dilution refrigerator (wet). To avoid heating effect from the magnetic field sweep, the climbing rate of the magnet was limited below 10 mT/min. Crossing zero-field heating effect was specially checked and avoided. We also made double checks by performing magnetic field dependent measurements in different manners to make sure that the field heating problem was avoided.

\subsection{\label{sec:level3} Time-dependent Ginzburg-Landau simulation}
To simulate the vortex state in a confined superconductor lead, the time-dependent Ginzburg–Landau (TDGL) equations are solved numerically. The TDGL equations are the time-dependent generalization of the Ginzburg-Landau (GL) equations [Tinkham, M. Introduction to superconductivity. Dover Public. Inc., 1996], which is a phenomenological theory for superconductors and, particularly, for a type-II superconductor in a vortex phase. The TDGL equations are a set of time-dependent nonlinear equations for the superconductor order parameter $\Psi$ and electromagnetic potentials of $\Phi$ and $\mathbf{A}$, which are written as,

\begin{equation} \begin{aligned}
\label{eqTDGLDimension}
\frac{\hbar^2\Psi}{2mD}(\frac{\partial}{\partial t}+i\frac{q}{\hbar}\Phi)
= -\frac{\Psi}{2m}(\frac{\hbar}{i}\nabla-q\mathbf{A})^2+\alpha\nonumber-\beta|\Psi|^2,
\end{aligned} \end{equation}

\begin{equation} \begin{aligned}
\sigma(\frac{\partial \mathbf{A}}{\partial t}+\nabla\Phi)
 =  \frac{q\hbar}{2mi}(\Psi^*\nabla\Psi-\Psi\nabla\Psi^*)-\frac{q^2}{m}|\Psi|^2 \mathbf{A}\\-\frac{1}{\mu_0}\nabla\times(\nabla\times \mathbf{A}
-\mathbf{B}),
\end{aligned} \end{equation}
where $\hbar$ is the reduced Planck's constant, $m$ is the Cooper pair mass, and $q=2e$ is the charge of a Cooper pair ($e$ is the charge of an electron). $\alpha$ and $\beta$ are two phenomenological parameters of the Gibbs free energy in the GL theory. Here, $\alpha$ is a temperature related parameter $\alpha(T)=\alpha(0)(1-T/T_c)$, with $T_c$ being the critical temperature of the superconductor. $\beta$ is a constant. $D$ and $\sigma$ are the diffusion coefficient and the conductivity in the superconductor, $\mu_0$ is the permeability of the free space [Alstrom et al. \emph{Appl. Math.} \bt{115}, 63 (2011); Gulian, A. Shortcut to superconductivity, Springer, 2020], and
$\mathbf{B}=B \mathbf{e}_z$ is the external magnetic field. The London penetration depth is $\lambda=\sqrt{m\beta/q^2 \mu_0 \alpha}$ and the GL coherence length is $\xi=\hbar/\sqrt{2m\alpha}$. The GL parameter is defined as the ratio of these two lengths $\kappa=\lambda/\xi$ with $\kappa<1/\sqrt{2}$ signaling a type-I superconductor and $\kappa>1/\sqrt{2}$ a type-II superconductor.

The dimensionless coordinate and time are scaled by $\lambda$ and $\xi^2/D$. Other dimensionless quantities can be defined in the corresponding units of
\begin{equation} \begin{aligned}
  \Psi_0 &=\sqrt{\alpha/\beta},             &
  \sigma_0 &=1/\mu_0 D \kappa^2,            &
  \Phi_0 &=\hbar D \kappa/q\xi^2,\\
  A_0 &=\hbar/q\xi,                &
  B_{0} &=\hbar/q \kappa \xi^2.   &
\end{aligned} \end{equation}
as,
\begin{equation} \begin{aligned}
  \psi &=\Psi/\Psi_0,~             &
  \sigma' &=\sigma/\sigma_0,       &
  \phi &=\Phi/\Phi_0,\\
  \mathbf{A}' &=\mathbf{A}/A_0,   &
  \mathbf{b} &=\mathbf{B}/B_{0}. &
\end{aligned} \end{equation}

After algebraic calculations and a gauge transformation, a dimensionless form of the TDGL equations can be obtained as,
\begin{equation} \begin{aligned}
\partial_\tau \psi  =  -(\frac{i}{\kappa}\nabla+\mathbf{A})^2\psi+\psi-|\psi|^2\psi,
\end{aligned} \end{equation}
\begin{equation} \begin{aligned}
\sigma \partial_\tau \mathbf{A}  =  \frac{1}{2i\kappa}(\psi^*\nabla\psi-\psi\nabla\psi^*)-|\psi|^2 \mathbf{A}-\nabla\times(\nabla\times \mathbf{A}-\mathbf{b}),
\end{aligned} \end{equation}
where $\sigma'$ and $\mathbf{A}'$ are relabeled as $\sigma$ and $\mathbf{A}$.

An extended boundary condition [Gulian, A. Shortcut to superconductivity, Springer, 2020] is adopted, which is done by attaching the superconductor with a non-superconducting, insulating material. Mathematically, the boundary conditions become
\begin{eqnarray}
\psi|_{\partial \Omega} & = & 0,\\
\nabla\times \mathbf{A}|_{\partial \Omega} & = & \mathbf{b}|_{\partial \Omega},
\end{eqnarray}
where $\partial \Omega$ represents the superconductor boundary. The initial condition for the TDGL equations is $ \psi|_{\tau=0}=1 $ in the superconductor region. With the boundary and initial condition, all the quantities, $\psi=\psi(t)$ and $\mathbf{A}=[A_x (t),A_y (t)]$ can be solved via the TDGL equations.

The dimensionless TDGL equations are then applied to the superconductor leads with two geometries. All results are computed in a dimensionless form. The leads are modeled as rectangles with width $w_1 \approx 10$ and the length $h_1 \approx 2.67$ for geometry-I (corresponding to Devices 1 and 2) and with $w_2 \approx 6.2$ and $h_2 \approx 1.17$ for geometry-II (corresponding to Device 3) in units of the London penetration depth $\lambda=150$ nm. We explore the parameter region of the superconductor with the GL parameter set at $\kappa =2$ and the dimensionless conductivity chosen as $\sigma=1$.

The external magnetic field is swept from 0 to the value at which the superconductor transits to the vortex state. For each magnetic field, the simulation of the Cooper pair density $\rho=|\psi|^2$ is performed for sufficiently long time ($\tau_{End}=10^4$) to assure the resulting state is the equilibrium state. Finally the number of vortices is counted. The results are shown in Fig. \ref{Fig4} in the main text, which illustrates that the magnetic fields at which the superconductor transits to the vortex states are different for the superconductor with different geometrical sizes.

\subsection{\label{sec:level4} Extended data for Device~1}
\begin{figure}[h]
\centering
\includegraphics[width=8.5 cm]{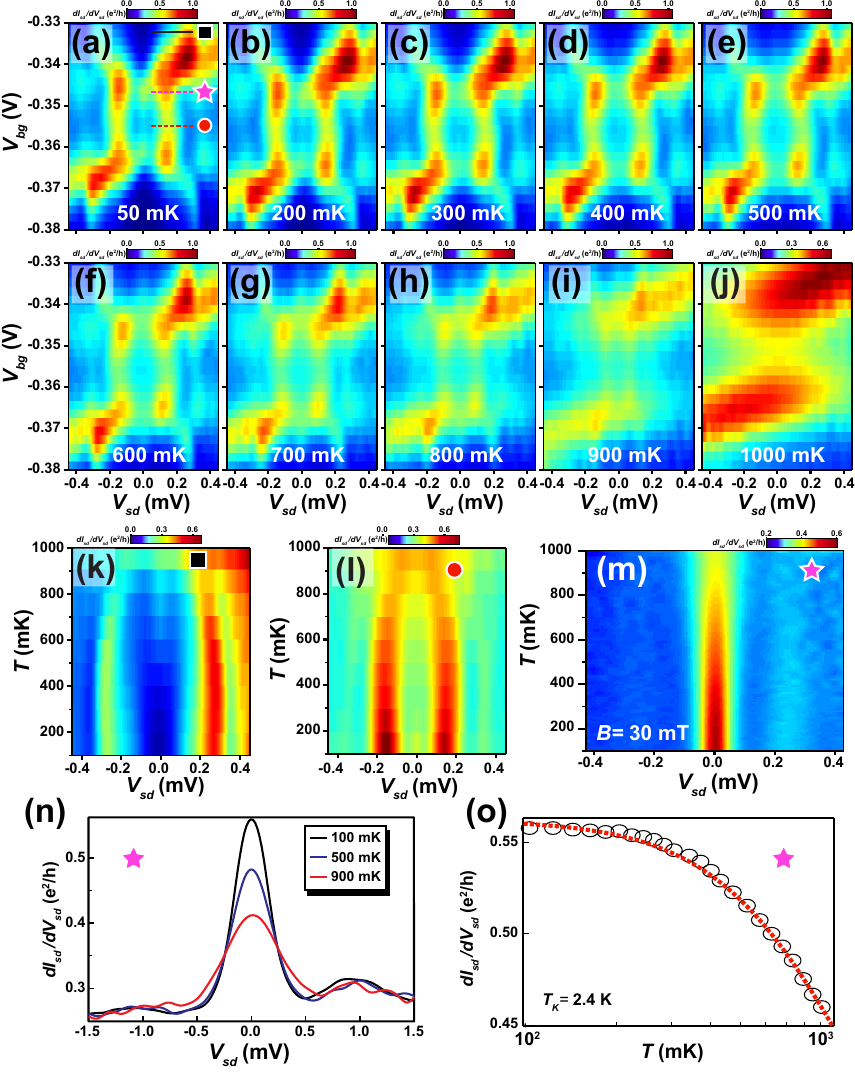}
\caption{\label{FigAPP_Dev1} Temperature dependence measurements in the K-II valley of Device~1.}
\end{figure}

Figure~\ref{FigAPP_Dev1} displays more measurements of Device~1. (a) Conductance spectrum in the K-II valley of the device in the superconducting state at $T=50$~mK and $B=0$ T. (b)-(j) Similar to (a), but at temperature from $T=200$~mK to $T=1000$~mK. (k)-(l) Extracted dataset as a function of temperature at given back gate voltages as indicated in (a). From panels (a)-(l), it can be seen that the superconductivity features are washed out gradually as temperature increases. Note that the emerged zero-bias conductance peak at high temperature in (k) can be attributed to thermal excitation of quasiparticles [Li, S. et al. \emph{Phys. Rev. B} \textbf{95}, 014515 (2017)]. (m) Temperature dependence of the Kondo peak measured for the device in normal state ($B=30$ mT) at a back gate voltage as indicated in (a). (n), Line-cuts taken from panel~(m). (o) Zero-bias conductance of the device in the normal state ($B=30$ mT) as a function of temperature (circles) and result of fitting (dashed line) according to the relation $G=G_{0}\left ( \frac{{{T}^{'}_K}^{2}}{T^{2}+{{T}^{'}_K}^{2}}\right )^s$, where ${T}^{'}_K=\frac{T_K}{{\left ( 2^{1/s}-1 \right )}^{1/2}}$ and $s=0.22$ [D. Goldhaber-Gordon. et al. \emph{Phys. Rev. Lett.} \textbf{81}, 5225 (1998)]. The fitting yields the Kondo temperature in the normal state $T_K\approx 2.4$~K.

\subsection{\label{sec:level5} Extended data for Device~2}
\begin{figure}[h]
\centering
\includegraphics[width=8 cm]{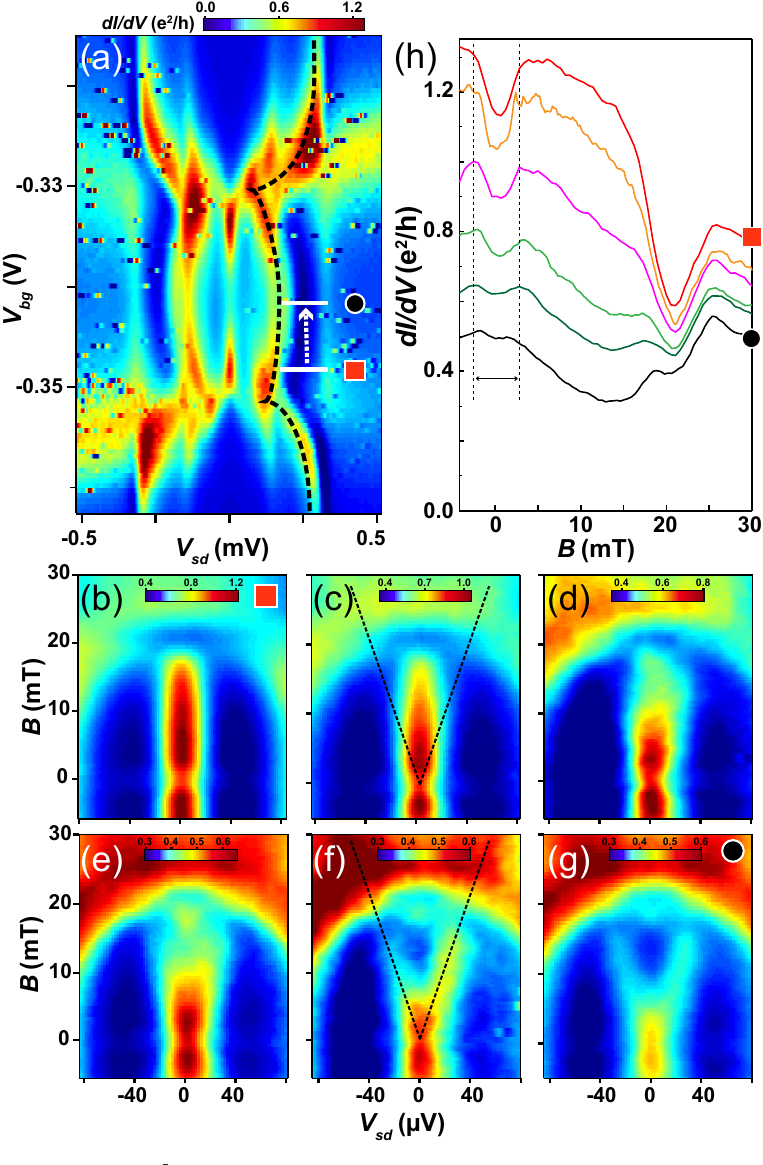}
\caption{Gate voltage dependence of NMRs. }
\label{FigAPP_Dev2_1}
\end{figure}

Figure~\ref{FigAPP_Dev2_1} shows the gate voltage dependence measurement of the NMR for Device~2. (a) Conductance spectrum in a Kondo regime. From the curvature of the ABS conductance peak (dashed line), it is evident to tell that the quantum dot is in a ``$\pi$"-junction configuration [Kim, B.-K. \emph{et al.} \emph{Phys. Rev. Lett.} \textbf{110}, 076803 (2013)]. (b)-(g), Magnetic field evolutions of the conductance spectra measured at different back gate voltages covering from the resonance region [panel (b)] to the deep Coulomb blockade region [(panel (g)]. (h) Magnetic field dependence of the zero-bias conductance, taken from panels (b)-(g), respectively. It can be seen clearly that the NMR occurs at all the gate voltages with ZBPs, and the NMR width is approximately the same for all $V_{\rm bg}$. 

Since $T_K$ strongly depends on the gate voltage through the relation $T_K\propto \frac{1}{2}\sqrt{\Gamma U}e^{\pi\epsilon (\epsilon+U)/\Gamma U}$ [Haldane, F. D. M. \emph{Phys. Rev. B} \textbf{40}, 416 (1978)], the NMR width thus hardly depends $T_K$. All data are taken from Device~2 with $V_{\rm g1}=-2.4$~V and $V_{\rm g2}=1.2$~V, and the values of $U = 2.9$ meV, $g^* \approx63$, and $\Delta\approx0.15$ meV are extracted for the device in the measured Kondo-correlated Coulomb blockade region.

\begin{figure}[h]
\centering
\includegraphics[width=8.5 cm]{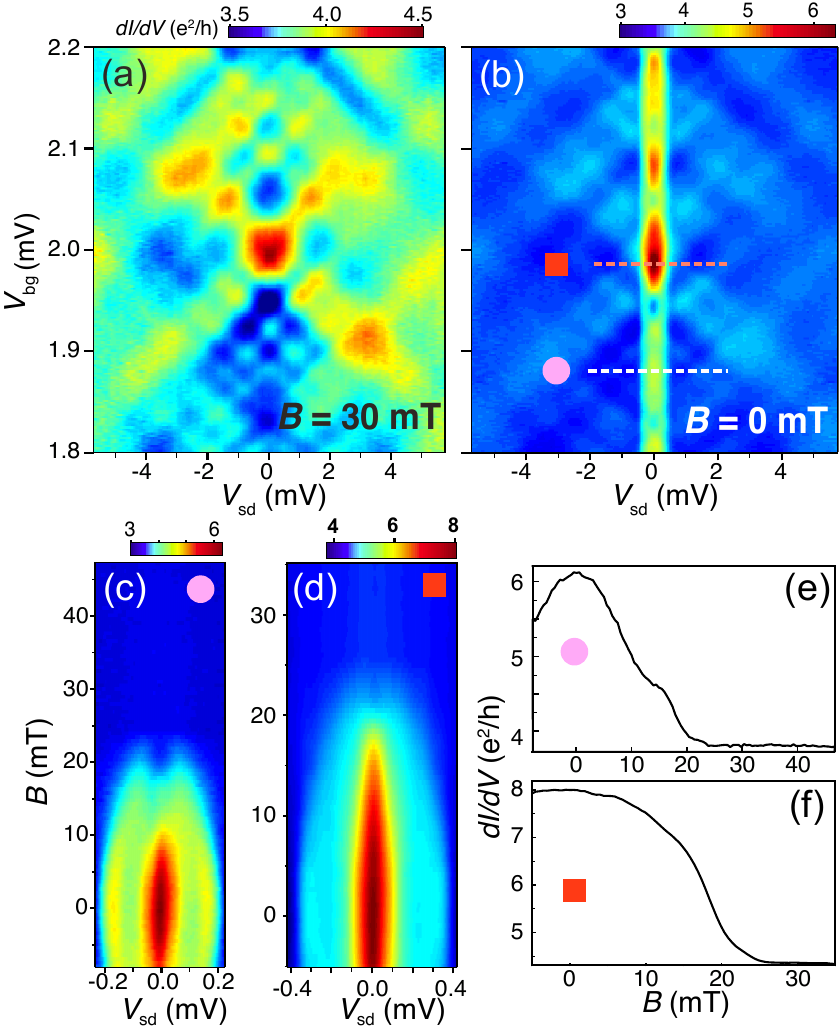}
\caption{Measurements of the conductance spectra for Device~2 in the open regime.}
\label{FigAPP_Dev2_2}
\end{figure}

Figure~\ref{FigAPP_Dev2_2} characterizes the open regime of Device~2, where no Coulomb blockade is present. (a) Differential conductance as a function of $V_{sd}$ and $V_{\rm bg}$ measured for Device~2 in the open regime at $B=30$~mT and $V_{\rm g1}=V_{\rm g2}=0$. A Fabry-P\'{e}rot interference conductance chess-board pattern can be identified, which shows that the device is in the open ballistic transport regime. (b) The same as in panel~(a), but the measurements are taken at $B=0$. The Fabry-P\'{e}rot pattern remains, but it is greatly modulated by the superconductivity in the low bias voltage region [Li. et al. \emph{Sci. Rep.} \textbf{6}, 24822 (2016)]. (c)  Measured differential conductance as a function of $V_{\rm sd}$ and $B$ at $V_{\rm bg}$= 1.88 V as indicated by the white dashed line in panel~(b). Supercurrent-induced zero-bias peak and finite-bias MAR peaks are observed to gradually vanish as the magnetic field increases. (d) Similar to panel~(c), but measured at the back gate voltage as indicated by the red dashed line in panel-(b), i.e., in an even more transparent region. (e)-(f) Line-cut plots at $V_{\rm sd}=0$ taken from panels~(c)-(d). Here, the zero-bias conductance is seen to  monotonically decrease with increasing magnetic field and no NMR is observed.

\subsection{\label{sec:level6} Extended data for Device~3}

\begin{figure}[h]
\centering
\includegraphics[width=8.5 cm]{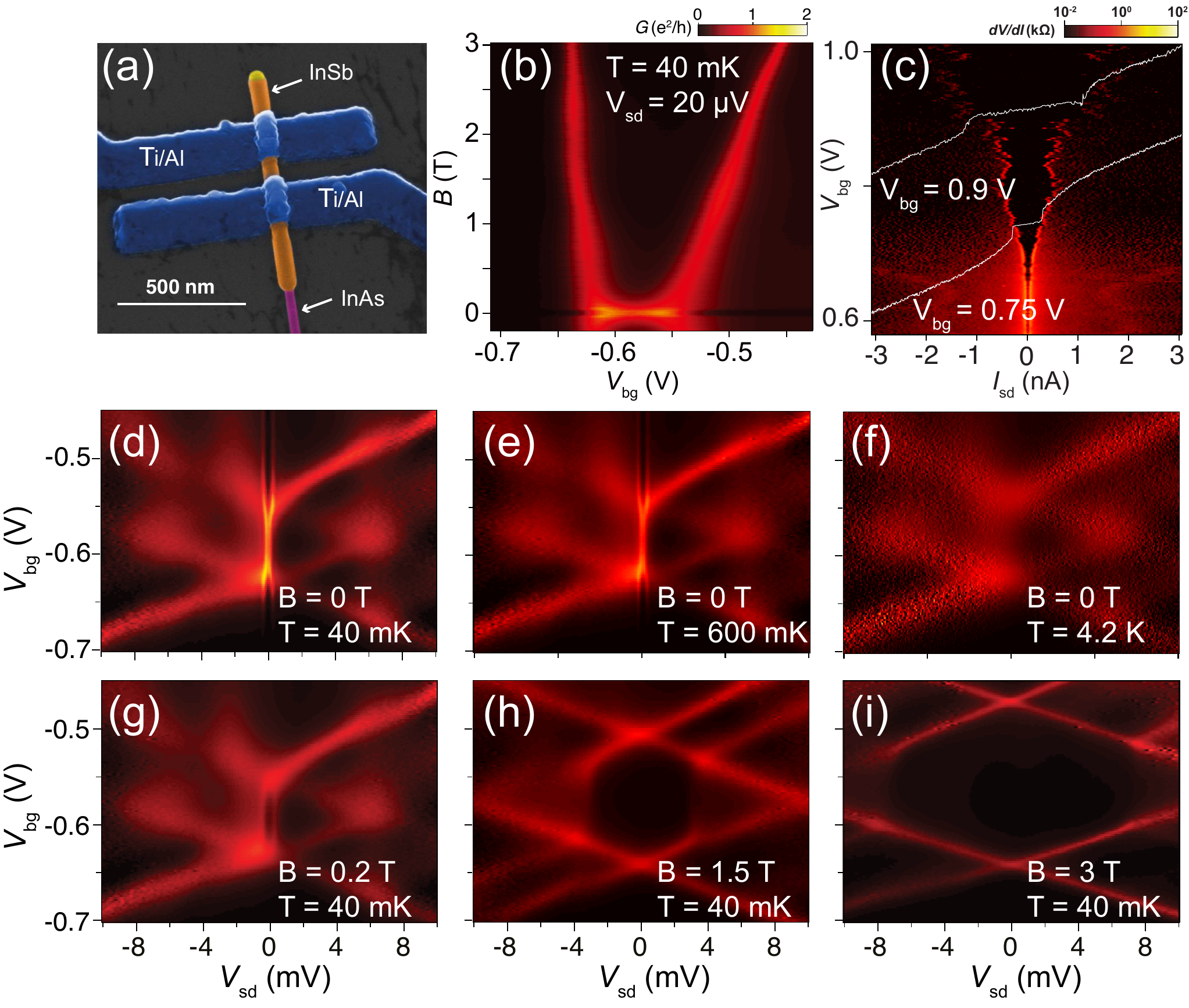}
\caption{Basic characterization of a Kondo valley in Device~3.}
\label{FigAPP_Dev3_1}
\end{figure}

Figure~\ref{FigAPP_Dev3_1} shows basic characterizations of Device~3. (a) SEM image of Device~3. Note the contacts are much slimmer than Devices~1 and 2. (b) Conductance as a function of magnetic field and back gate voltage in the linear response regime. The quantum level $g$-factor in the dot can be extracted from the level splitting in the magnetic field, which yields $g\approx 40$. It can be also extracted by the Kondo peak splitting. (c) Differential resistance as a function of source-drain current and back gate voltage measured in an open conduction region. The middle dark region corresponds to dissipationless superconducting region and the critical current shows a strong gate voltage dependence. The white lines are representative results of measurements for the source-drain voltage as a function of the applied source-drain current at given back gate voltages. The supercurrent decreases fast as the device goes to Coulomb blockade region, and the Cooper pair tunneling has to be measured by the conductance peak. (d)-(f) Conductance spectra measured at different temperatures. The Kondo peak is gradually washed out here, but it can remains when the temperature beyond the critical temperature of the contacts. (g)-(h) Conductance spectra measured at different magnetic fields. The cotunneling conductance shows the splitting of the Kondo effect.

\begin{figure}[h]
\centering
\includegraphics[width=8.5cm]{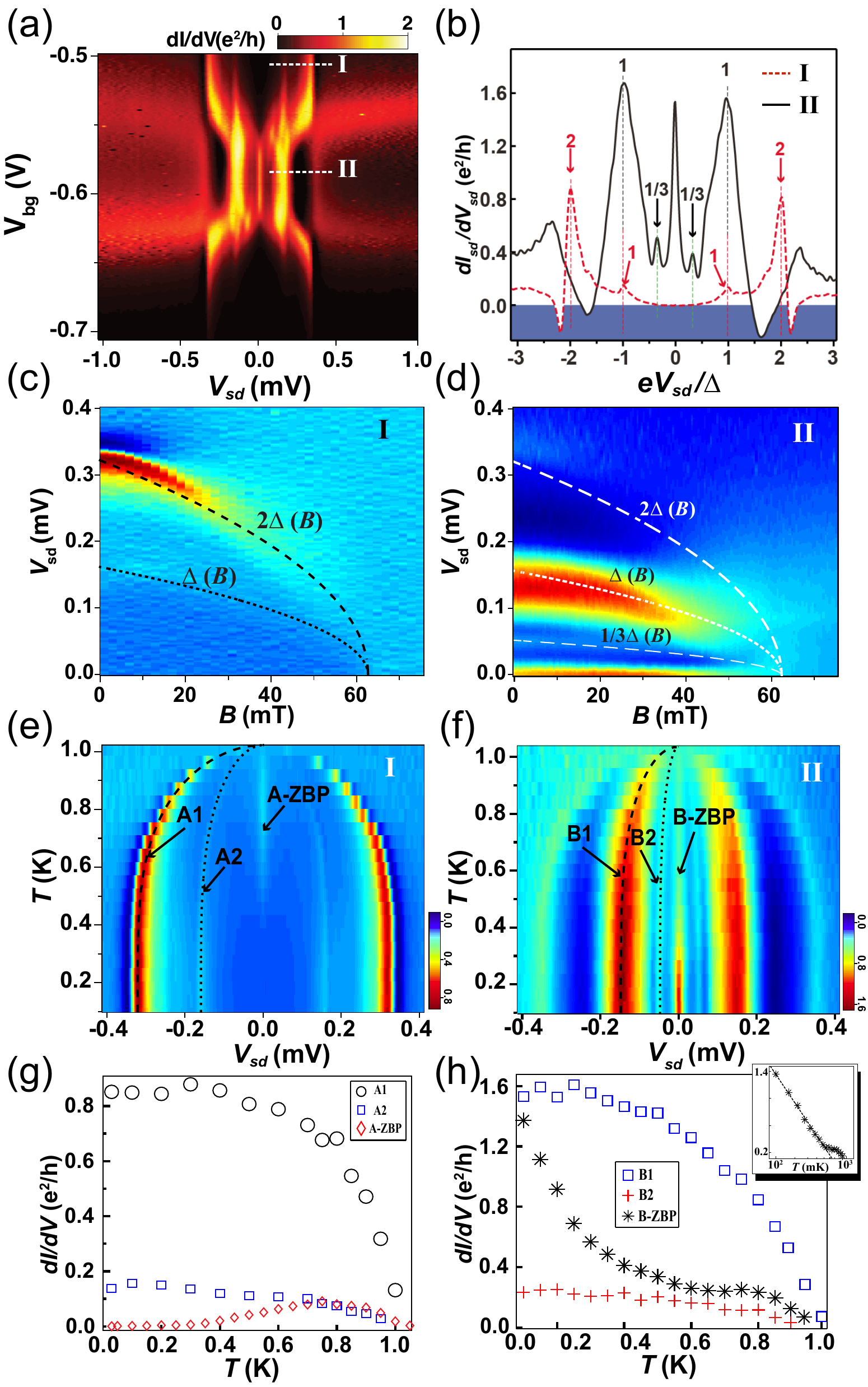}
\caption{Basic characterization of sub-gap structures in Device~3. }
\label{FigAPP_Dev3_2}
\end{figure}

Figure~\ref{FigAPP_Dev3_2} focuses the sub-gap conductance characterization of Device~3. (a) Low energy conductance spectral measurements of the device in the superconducting state. (b) Line-cuts taken from panel~(a) at gate voltages as marked by white dashed lines in panel~(a). Multiple Andreev reflection (MAR) peaks of different orders are seen in the conductance plots. (c) and (d) Magnetic field evolutions of the sub-gap structures at the gate voltage marked by white dashed line ``I" in panel~(a), corresponding to an even-number electron occupation in the dot, and at the gate voltage marked by white dashed line ``II" in panel~(a) corresponding to an odd-number electron occupation in the dot. The positions of the MAR-peaks in both cases follow the predictions of the BCS theory in which the superconducting gap evolves with the magnetic field as $\Delta \left ( B \right )=\Delta _{0}\sqrt{1-\left (  \frac{B}{B_{c}}\right )^{2}}$. (e) and (f) Temperature evolutions of the sub-gap structures at the same back gate voltages as in panels~(c) and (d). Again, the positions of the MAR-peaks follow the relation: $\Delta \left (T\right )=\Delta _{0}\sqrt{\cos \left [ \frac{\pi }{2} \left ( \frac{T}{T_{c}} \right )^{2} \right]}$. (g) and (h) Temperature evolutions  of the peak height extracted from panels~(e) and (f).

It can be seen that the heights of the MAR-peaks evolve with temperature very differently from the Josephson current-induced zero-bias peak (``B-ZBP"). The inset of panel-(h) shows the plot of the ``B-ZBP" peak in panel-(f) as a function of temperature on a logarithmic scale, which illustrates a typical temperature dependence of a Kondo peak [Cronenwett. et al. \emph{Science} \textbf{281}, 540 (1998)]. This is consistent with the fact that the Josephson current strongly depends on the Kondo correlation present here. Again, the zero-bias peak emergent at high temperature (``A-ZBP") in panel~(g) can be ascribed to thermal excitation of quasiparticles. In fact, the thermal excitation also causes a little bump on the ``B-ZBP" peak in the same temperature range.

\end{appendix}


\begin{thebibliography}{99}

\bibitem{Gennes1964}
P.D.~Gennes and M.~Tinkham, \textit{Magnetic behavior of very small superconducting particles. Vol. 1} (Universite de Paris, 1964). 

\bibitem{Gennes1989}
P.D.~Gennes, \textit{Superconductivity of metals and alloys} (Addison-Wesley, New York, 1989).

\bibitem{Jaccarino1962}
V.~Jaccarino and M.~Peter, Phys. Rev. Lett. \textbf{9}, 290 (1962).

\bibitem{Maekawa1978}
S.~Maekawa and M.~Tachiki, Phys. Rev. B \textbf{18}, 4688 (1978).

\bibitem{Meul1984}
H.W.~Meul, C.~Rossel, M.~Decroux, \O.~Fischer, G.~Remenyi, and A.~Briggs, Phys. Rev. Lett. \textbf{53}, 497 (1984).

\bibitem{Uji2001}
S.~Uji, H.~Shinagawa, T.~Terashima, T.~Yakabe, Y.~Terai, M.~Tokumoto, A.~Kobayashi, H.~Tanaka, and H.~Kobayashi, Nature \textbf{410}, 908 (2001).

\bibitem{Kogan1987}
V.G.~Kogan and N.~Nakagawa, Phys. Rev. B \textbf{35}, 1700 (1987).

\bibitem{Abrikosov1969}
A.A.~Abrikosov, Sov. Phys. Uspekhi \textbf{12}, 168 (1969).

\bibitem{Maki1969}
K.~Maki, \textit{Superconductivity} (Marcel Dekker, New York, 1969).

\bibitem{Nilsson2009}
H.A.~Nilsson, P.~Caroff, C.~Thelander, M.~Larsson, J.B.~Wagner, L.-E.~Wernersson, L.~Samuelson, and H.Q.~Xu, Nano Lett. \bt{9}, 3151 (2009).

\bibitem{Kretinin2010}
A.V.~Kretinin, R.~Popovitz-Biro, D.~Mahalu, and H.~Shtrikman, Nano Lett. \bt{10}, 3439 (2010).

\bibitem{Gordon1998}
D.~Goldhaber-Gordon, J.~G\"{o}res, M.A.~Kastner, H.~Shtrikman, D.~Mahalu, and U.~Meirav, Phys. Rev. Lett. \bt{81}, 5225 (1998).

\bibitem{Kim2013}
B.-K.~Kim, Y.-H.~Ahn, J.-J.~Kim, M.-S.~Choi, M.-H.~Bae, K.~Kang, J.S.~Lim, R.~L\'{o}pez, and N.~Kim, Phys. Rev. Lett. \bt{110}, 076803 (2013).

\bibitem{Li2017}
S.~Li, N.~Kang, P.~Caroff, and H.Q.~Xu, Phys. Rev. B \bt{95}, 014515 (2017).

\bibitem{Buitelaar2002}
M.R.~Buitelaar, T.~Nussbaumer, and C.~Sch\"{o}nenberger, Phys. Rev. Lett. \bt{89}, 256801 (2002).

\bibitem{Buitelaar2003}
M.R.~Buitelaar, W.~Belzig, T.~Nussbaumer, B.~Babi\'{c}, C.~Bruder, and C. Sch\"{o}nenberger, Phys. Rev. Lett. \bt{91}, 057005 (2003).

\bibitem{Jespersen2007}
T.~Sand-Jespersen, J.~Paaske, B.M.~Andersen, K.~Grove-Rasmussen, H.I.~J\o rgensen, M.~Aagesen, C.B.~S\o rensen, P.E.~Lindelof, K.~Flensberg, and J.~Nyg\aa rd, Phys. Rev. Lett. \bt{99}, 126603 (2007).

\bibitem{Eichler2007}
A.~Eichler, M.~Weiss, S.~Oberholzer, C.~Sch\"{o}nenberger, A.~Levy Yeyati, J.C.~Cuevas, and A.~Martín-Rodero, Phys. Rev. Lett. \bt{99}, 126602 (2007).

\bibitem{Pillet2010}
J.-D.~Pillet, C.H.~L.~Quay, P.~Morfin, C.~Bena, A. Levy Yeyati, and P.~Joyez, Nat. Phys. \bt{6}, 965 (2010).

\bibitem{Lee2014}
E.J.H.~Lee, X.~Jiang, M.~Houzet, R.~Aguado, C.M.~Lieber, and S.~De Franceschi, Nat. Nanotechnol. \bt{9}, 79 (2014).

\bibitem{Rasmussen2007}
K.~Grove-Rasmussen, H.I.~J\o rgensen, and P.E.~Lindelof, New J. Phys. \bt{9}, 124 (2007).

\bibitem{Lee2012}
E.J.H.~Lee, X.~Jiang, R.~Aguado, G.~Katsaros, C.M.~Lieber, and S.~De Franceschi, Phys. Rev. Lett. \bt{109}, 186802 (2012).

\bibitem{Mourik2012}
V.~Mourik, K.~Zuo, S.M.~Frolov, S.R.~Plissard, E.P. A. M.~Bakkers, and L.P.~Kouwenhoven, Science \bt{336}, 1003 (2012).

\bibitem{Deng2012}
M.T.~Deng, C.L.~Yu, G.Y.~Huang, M.~Larsson, P.~Caroff, and H.Q.~Xu, Nano Lett. \bt{12}, 6414 (2012).

\bibitem{Deng2016}
M.T.~Deng, S.~Vaitiek\.{e}nas, E.B.~Hansen, J.~Danon, M.~Leijnse, K.~Flensberg, J.~Nyg\aa rd, P.~Krogstrup, and C.M.~Marcus, Science \bt{354}, 1557 (2016).

\bibitem{Tiira2017}
J.~Tiira, E.~Strambini, M.~Amado, S.~Roddaro, P.~San-Jose, R.~Aguado, F.S.~Bergeret, D.~Ercolani, L.~Sorba, and F.~Giazotto, Nat. Commun. \bt{8}, 14984 (2017).

\bibitem{Fu2021}
J.-B.~Fu, B.~Li, X.-F.~Zhang, G.-Z~Yu, G.-Y.~Huang, and M.-T~Deng. Sci. China Physics, Mech. Astron. \bt{64}, (2021).

\bibitem{Cao2023}
Z. Cao, S. Chen, G. Zhang, and D.E. Liu. Sci. China Physics, Mech. Astron. \bt{66}, 267003 (2023).

\bibitem{Pikulin2012}
D.I.~Pikulin, J.P.~Dahlhaus, M.~Wimmer, H.~Schomerus, and C.W.J.~Beenakker, New J. Phys. \bt{14}, 125011 (2012).

\bibitem{vandam2006}
J.A.~van Dam, Y.V.~Nazarov, E.P.A.M.~Bakkers, S.~De Franceschi, and L.P.~Kouwenhoven, Nature \bt{442}, 667 (2006).

\bibitem{Chang2013}
W.~Chang, V.E.~Manucharyan, T.S.~Jespersen, J.~Nyg\aa rd, and C.M.~Marcus, Phys. Rev. Lett. \bt{110}, 217005 (2013).

\bibitem{Lee2022}
M.~Lee, R.~L\'{o}pez, H.Q.~Xu, and G.~Platero, Phys. Rev. Lett. \bt{129}, 207701 (2022).

\bibitem{Wang2014NC}
C.~Wang, Y.Y.~Gao, I.M.~Pop, U.~Vool, C.~Axline, T.~Brecht, R.W.~Heeres, L.~Frunzio, M.H.~Devoret, G.~Catelani, L.I.~Glazman, and R.J.~Schoelkopf, Nat. Commun. \bt{5}, 5836 (2014).

\bibitem{Ambegaokar1969}
V.~Ambegaokar and B.I.~Halperin, Phys. Rev. Lett. \bt{22}, 1364 (1969).

\bibitem{Pekola2013RMP}
J.P.~Pekola, O.P.~Saira, V.F.~Maisi, A. Kemppinen, M.~M\"{o}tt\"{o}nen, Y.A.~Pashkin, and D.V.~Averin, Rev. Mod. Phys. \bt{85}, 1421 (2013).

\bibitem{Stan2004PRL}
G.~Stan, S.B.~Field, and J.M.~Martinis, Phys. Rev. Lett. \bt{92}, 097003 (2004).

\bibitem{Sato2022PRL}
Y.~Sato, K.~Ueda, Y.~Takeshige, H.~Kamata, K.~Li, L.~Samuelson, H.Q.~Xu, S.~Matsuo, and S.~Tarucha, Phys. Rev. Lett. \bt{128}, 207001 (2022).

\bibitem{Vaitiekenas2018PRL}
S.~Vaitiek\.{e}nas, M.-T.~Deng, J.~Nyg\aa rd, P.~Krogstrup, and C.M.~Marcus, Phys. Rev. Lett. \textbf{121}, 037703 (2018).

\bibitem{Mikkelsen2018PRX}
A.E.G.~Mikkelsen, P.~Kotetes, P.~Krogstrup, and K.~Flensberg, Phys. Rev. X \textbf{8}, 031040 (2018).

\bibitem{Antipov2018PRX}
A.E.~Antipov, A.~Bargerbos, G.W.~Winkler, B.~Bauer, E.~Rossi, and R.M.~Lutchyn, Phys. Rev. X \bt{8}, 031041 (2018).

\bibitem{Loo2023}
N.~van Loo, G.P.~Mazur, T.~Dvir, G.~Wang, R.C.~Dekker, J.-Y.~Wang, M.~Lemang, C.~Sfiligoj, A.~Bordin, D.~van Driel, G.~Badawy, S.~Gazibegovic, E.P.A.M.~Bakkers, and L.P.~Kouwenhoven, Nat. Commun. \bt{14}, 3325 (2023).

\end{thebibliography}
\end{document}